\documentclass{article}
\usepackage{spconf,amsmath,graphicx}
\usepackage{setspace}
\usepackage{enumitem}
\setlist{nosep, leftmargin=14pt}
\usepackage{mwe} 
\usepackage{amssymb}
\usepackage[table,dvipsnames]{xcolor}
\usepackage{multirow}
\usepackage{bbm}
\usepackage[vlined]{algorithm2e}
\usepackage{subfigure}
\usepackage{subfloat}
\usepackage{float}
\usepackage{svg}
\definecolor{lightsalmon}{rgb}{1.0, 0.63, 0.48}

\newcommand{\support}{\mathbb{S}}
\newcommand{\query}{\mathbb{Q}}

\newcommand\bs[1]{\boldsymbol{#1}}

\title{A Transductive Few-Shot Learning Approach for Classification of digital histopathological slides from liver cancer}
\name{\begin{tabular}{@{}c@{}}
Aymen Sadraoui$^{1*}$ \thanks{$^*$ Equal contributions}\quad
S\'egol\`ene Martin$^{1*}$ \quad 
Eliott Barbot$^{1}$ \quad 
Astrid Laurent-Bellue$^{2,3,4}$\\  
Jean-Christophe Pesquet$^1$ \quad Catherine Guettier$^{2,3,4}$ \quad 
Ismail Ben Ayed$^5$ 
\end{tabular}}

\address{$^1$ Centre de Vision Num\'erique, Universit\'e Paris-Saclay, Inria, CentraleSup\'elec, Gif-sur-Yvette, France \\
$^2$Department of Pathology, AP-HP, H\^opital Bic\^etre, Le Kremlin-Bic\^etre, France\\
$^3$Facult\'e de M\'edecine, Universit\'e Paris-Saclay, Le Kremlin-Bic\^etre, France\\
$^4$INSERM U1193, Villejuif, France\\
$^5$ \'Ecole de Technologie Sup\'erieure de Montr\'eal, Montr\'eal, Canada \\
}
\begin{document}
%
\maketitle
\begin{abstract}
This paper presents a new approach for classifying 2D histopathology patches using few-shot learning. The method is designed to tackle a significant challenge in histopathology, which is the limited availability of labeled data. By applying a sliding window technique to histopathology slides, we illustrate the practical benefits of transductive learning (i.e., making joint predictions on patches) to achieve consistent and accurate classification.
Our approach involves an optimization-based strategy that actively penalizes the prediction of a large number of distinct classes within each window. We conducted experiments on histopathological data to classify tissue classes in digital slides of liver cancer, specifically hepatocellular carcinoma. The initial results show the effectiveness of our method and its potential to enhance the process of automated cancer diagnosis and treatment, all while reducing the time and effort required for expert annotation.

\end{abstract}

\section{Introduction}
\label{sec:intro}
In clinical settings, histopathology images are a critical primary source of information for pathologists to perform cancer diagnostics and choose treatment strategies. With the widespread adoption of digital pathology, it has become a standard practice to digitize histology slides into high-resolution images called Whole Slide Images (WSIs). WSIs have initiated a new era offering
considerable opportunities for using AI assistance systems \cite{neofytos2019,xiang2021overview}. In particular, supervised deep learning methods based on conventional neural networks (CNNs) have made great strides in cancer research \cite{shin2016deep,saillard2020predicting}. However, the success of classical supervised learning approaches depends on the availability of extensive annotated training data. Unlike natural images, which can be annotated via crowd-sourcing, histopathology necessitates expert pathologists' accurate annotations of gigapixel-sized images. Due to the time-consuming nature of the labeling process, histopathology datasets tend to be limited in size, which poses significant challenges for training machine learning models \cite{cooper2023machine}. Moreover, WSIs can exhibit variability due to staining techniques, tissue preparation, and image quality \cite{van2010interobserver}, affecting supervised model performance. Furthermore, supervised learning models can encounter difficulties when confronted with imbalanced data, a common scenario in histopathology. Non-uniform class distribution may produce biased results and compromise the model performance \cite{johnson2019survey}.
Few-shot learning methods address the limitations found in traditional supervised learning techniques, providing efficient models capable of generalizing from a small set of labeled examples.
These methods not only prove to be scalable, but also significantly reduce costs and time consumption.\\
Transductive few-shot learning \cite{martin2022towards,veilleux2021realistic}, a particularly appealing category within this field, has a distinct advantage. Unlike supervised classification methods, which often treat each data sample independently, transductive methods make predictions on a set of samples collectively. This is especially useful when dealing with localized regions in medical imaging. It allows us to leverage homogeneity and spatial coherence across multiple patches in such a  region to enhance the classification accuracy and reliability.\\
In this work, we introduce a novel transductive few-shot learning approach for histopathological image classification. To our knowledge, it is the first of this kind in the field \cite{szolomicka2023overview}. Our main contributions are
summarized below.
\begin{itemize}
\item We apply a sliding window technique to WSIs, establishing a practical scenario where the advantages of transductive few-shot learning are clearly demonstrated.
\item Inspired by previous work \cite{martin2022towards}, we develop an optimization-based method for few-shot classification of histopathological patches.
\item We validate our approach by tests on the most frequent liver cancer (i.e., hepatocellular carcinoma, HCC), showcasing its effectiveness and confirming its high potential for practical application.
\end{itemize}
The paper is organized as follows. Section \ref{sec:data} describes the medical context. The few-shot methodology and the proposed algorithm are detailed in Section \ref{sec:method}. Finally, experimental results are presented in Section \ref{sec:experiments}, and Section \ref{sec:conclu} is dedicated to the conclusion.

\section{Medical multiclass problem}\label{sec:data}
We used HCC WSIs stained with HES (Hemaloxylin-Eosin-Saffron) and digitized at 40$\times$ magnification. More precisely, we aim to classify local tissues into the following five classes:
 \begin{enumerate}
    \item Non-Tumor Liver (NT): Liver sections that are not affected by HCC but may be affected by cirrhosis.
    \item Hemorragic tissue (RE): a non-tumoral pattern characterized by blood cell suffusion.
    \item Tumor tissue with macro-trabecular architecture (AM): An aggressive pejorative tumor type characterized by trabeculae of more than ten cells thick.
    \item Tumor tissue with Vessels Encapsulating Tumor Clusters architecture (VE): An aggressive pejorative tumor type characterized by tumor cells arranged in small clusters and surrounded by endothelial cells.
    \item Conventional trabecular architecture (AN): A non-pejorative tumoral pattern commonly found in HCC patients.
    
\end{enumerate}
The distinction between tumor and non-tumor areas and the evaluation of pejorative tumor areas provide insightful information to medical doctors.\\
In this paper, 28 patients from a  previously formed cohort of 108 patients with HCC were selected from usable HES slides from Kremlin-Bic\^etre Hospital, France, 
and manually annotated by two skilled pathologists in the five above categories. Annotated WSIs were then tiled into $1728\times 1728$ patches. Data distribution per class is displayed in Table \ref{table:data_dist}.
\vspace{-0.1cm}
\begin{table}[ht]
\begin{center}
\begin{tabular}{|l|l|l|l|l|l|}
\hline
Class & NT & RE & AM & VE & AN\\ \hline
Percentage & 26\% & 14\% & 8\% & 12\% & 40\% \\ \hline
\end{tabular}
\vspace{-0.1cm}
\caption{Data distribution per class.}
\label{table:data_dist}
\end{center}
\end{table}
\vspace{-0.5cm}

\section{Proposed method}\label{sec:method}
\subsection{Problem formulation}
Few-shot methods typically involve a two-step process \cite{chen2019closer}: first, a neural network, pre-trained on a comprehensive and generic dataset, extracts features from the images of interest. Then, a specifically designed classifier is applied to these extracted features to perform the classification task.\\
We start by introducing the notation for the few-shot classification challenge at hand. 
The pre-trained network encoder, denoted by $\Phi$, is crucial for feature extraction. Typically, it has been trained on a dataset $\mathcal{D}_{\text{base}}$ encompassing a broad spectrum of images, potentially inclusive of various WSIs from a multitude of organs and medical facilities. Still, it may not precisely encapsulate the exact categories of our specific classification tasks.\\
The few-shot dataset consists of $N$ images spanning across $K$ distinct classes. 
In our context of few-shot classification for histopathological images, $K$ equals 5.
Within the dataset, a subset, referred to as the \emph{support set} with index set $\support \subseteq \{1, \dots, N\}$, encompasses the feature samples $(\boldsymbol{x}_n)_{n \in \support}$ and their respective one-hot-encoded labels $(\boldsymbol{y}_n)_{n \in \support}$.\footnote{For every $n \in \support$ and $k \in \{1, \dots, K\}$, $y_{n, k} = 1$ if $\boldsymbol{x}_n$ is an instance of class $k$, and $y_{n, k} = 0$ otherwise.} The support set is constituted by $s$-\emph{shots} (labeled examples) for each class. In contrast, the \emph{query set} with indices in $\query = \{1, \dots, N\} \setminus \support$, comprises a batch of unlabeled samples $(\boldsymbol{x}_n)_{n \in \query}$.
The goal is to accurately predict the labels for the samples of the query set under the supervision of the support set. To achieve this, the representations 
$(\boldsymbol{z}_n = \Phi(\boldsymbol{x}_n))_{1\le n \le N}$
generated by the feature extractor are fed into our few-shot classifier.

\subsection{Transductive methodology}
\begin{figure}[ht]
    \centering
    \includegraphics[width=0.55\linewidth]{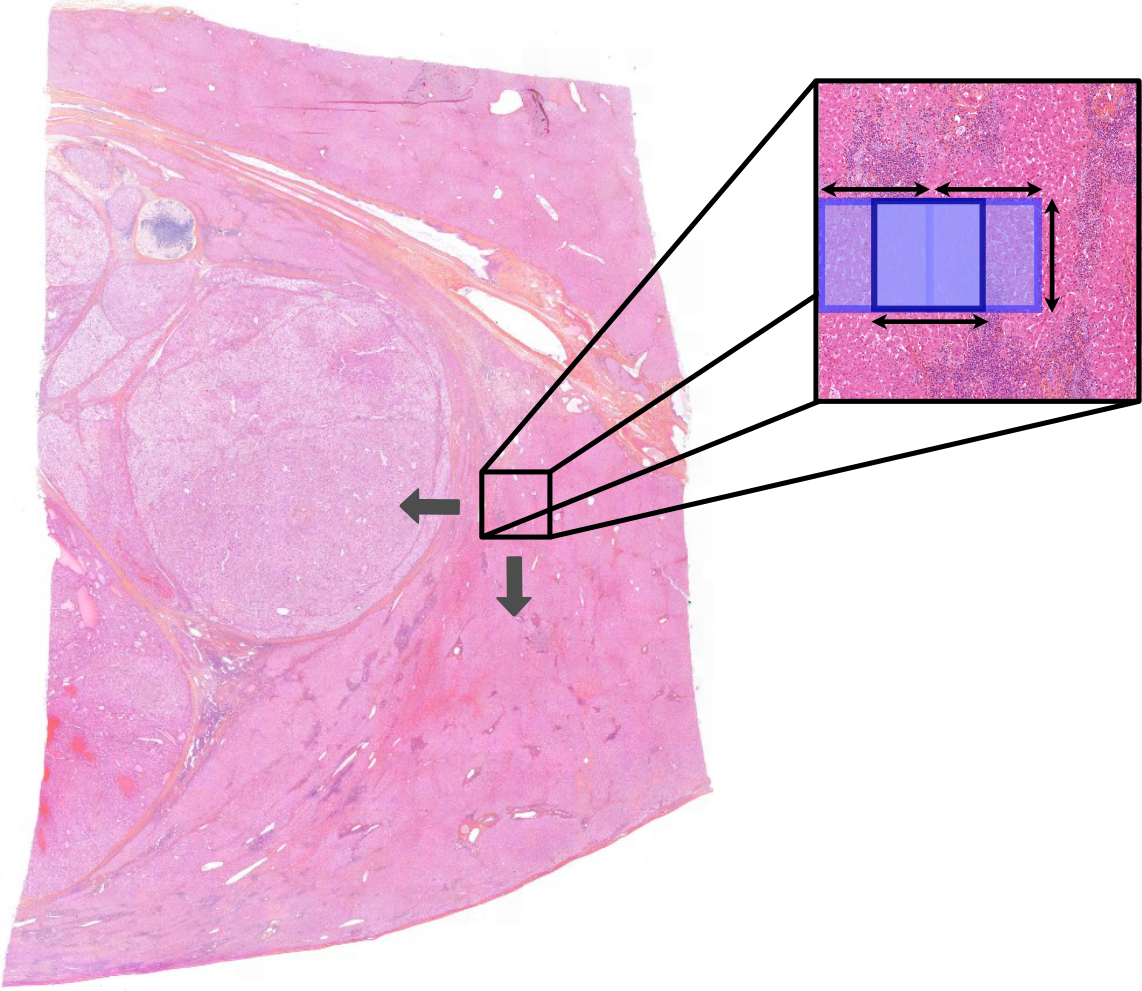}
    \caption{Scanning of the slide with a sliding window.}
    \label{fig:sliding_window}
\end{figure}

One of the primary advantages of few-shot learning methodologies, when contrasted with traditional supervised learning techniques, lies in their ability to collectively infer from an entire batch of $|\query|> 1$ query instances simultaneously rather than evaluating each instance independently. In the lexicon of few-shot learning, this methodology is referred to as \emph{transductive} learning \cite{bronskill2020tasknorm,hu2021leveraging}.
Transductive few-shot methods are designed to make joint predictions for the entire batch of query samples within each specific few-shot task. This approach takes full advantage of the statistical properties inherent to the query set of a task, employing shared information across instances to enhance generalization and accuracy. Empirical studies have demonstrated that batch-based inference on unlabeled instances, as opposed to individual sample evaluation, results in substantial improvements in prediction accuracy \cite{joachims1999transductive}.\\
In the field of microscopy analysis, where spatial pattern recognition is crucial, transductive few-shot approaches exhibit significant potential. Commonly, in a single WSI, it is observed that architectures belonging to the same class tend to cluster spatially, forming homogeneous regions. To leverage this spatial coherence, our strategy involves selecting a window of dimensions $S \times S$ on the microscope slide, as depicted in Figure \ref{fig:sliding_window}. Each window comprises overlapping mini-patches, each of dimensions $s \times s$, constituting the query set for our few-shot task. The underlying assumption here is that each window encapsulates a few (typically, one or two) distinct classes, allowing each mini-patch to serve as an additional (unlabeled) instance of these classes. By sliding the window across the entire WSI, we facilitate comprehensive predictions across its entirety.

\subsection{Minimization problem}
Our method estimates the optimal class assignments of each small patch within the window while limiting the number of predicted classes, thereby acknowledging the spatial coherence in such samples.\\
By refining the method presented in \cite{martin2022towards}, we approach the few-shot classification challenge through a minimization problem, seeking optimal solutions for the one-hot-encoded assignments $\boldsymbol{U} = (\boldsymbol{u}_n)_{1 \leq n \leq |\query|} \in (\Delta_K)^{|\query|}$ and the class centroids $\boldsymbol{W} = (\boldsymbol{w}_k)_{1 \leq k \leq K} \in (\mathbb{R}^d)^K$, where $\Delta_K$ represents the unit simplex set in $\mathbb{R}^K$. The problem is mathematically formulated as
\begin{alignat}{2}\label{eq:minimization_pb_histo}
    &\underset{\boldsymbol{U}, \boldsymbol{W}}{\text{minimize}}\,   && f(\boldsymbol{U}, \boldsymbol{W}) + g(\boldsymbol{U}) + \lambda\, h(\boldsymbol{U}) ,\\
    &\text{subject to} \quad &&  (\forall n \in \query) \quad \boldsymbol{u}_n \in \Delta_K, \nonumber\\
    & && (\forall n \in \support)\quad\; \boldsymbol{u}_{n} = y_{n},\nonumber
\end{alignat}
with $\lambda$ is a positive regularization parameter.
Here, $f$ represents the data-fidelity term, reflecting the assumption that the data follows a multivariate Gaussian distribution and integrating supervision from the support set. Formally, we define
\begin{multline}
     f(\boldsymbol{U}, \boldsymbol{W}) = \frac{1}{2}\sum_{k=1}^K \sum_{n=1}^N  u_{n, k}  (\boldsymbol{w}_k - \boldsymbol{z}_n)^\top \hat{\bs{S_k}}(\boldsymbol{w}_k - \boldsymbol{z}_n)\\ -\frac{1}{2} \sum_{k=1}^K \sum_{n=1}^N u_{n,k}\ln \det(\hat{\bs{S_k}})
\end{multline}
where, for every $k \in \{1, \dots, K\}$, $\hat{\bs{S_k}}$
is a symmetric positive matrix corresponding to
a sparse approximation of inverse of the empirical covariance matrix of class $k$, computed from the support set with a Graphical Lasso approach \cite{friedman2008sparse}.
In addition, 
$g$ represents an entropic barrier on the assignments, facilitating closed-form updates in the forthcoming algorithm. It is expressed as
\begin{equation}
    g(\boldsymbol{U}) = \sum_{k=1}^K \sum_{n\in \query} u_{n,k} \ln u_{n,k}.
\end{equation}
Finally, the penalty function $h$ is central to our approach: it acts as a partition complexity term, encouraging a minimal number of classes to be predicted within the window:
\begin{equation}
    h(\boldsymbol{U})  = - \sum_{k=1}^K \pi_{k} \ln(\pi_{k}),
\end{equation}
where, for every $k \in \{1, \dots, K\}$, $\pi_k = \frac{1}{|\query|}\sum_{n \in \query} u_{n, k}$ denotes the proportion of samples of class $k$ in the query set.

\subsection{Algorithm}
To address the minimization problem outlined in Equation~\eqref{eq:minimization_pb_histo}, we propose an algorithm that alternates minimization steps with respect to the variables $\boldsymbol{U}$ and $\boldsymbol{W}$. Our iterative approach, detailed in Algorithm \ref{algo:PADDLE}, shares similarities with the technique presented in \cite{martin2022towards}, the primary distinction being the introduction of inverse covariance matrices. Given these similarities, we 
direct the reader to \cite{martin2022towards} for 
more details on our methodology and
the convergence guarantees of the algorithm.

\begin{center}
\RestyleAlgo{ruled}
	\begin{algorithm} 
\DontPrintSemicolon
Initialize $\boldsymbol{W}^{(0)}$ as the means computed on the support set and for all $k\in \{1, \dots, K\}$,  $\displaystyle\pi_{k}^{(0)} = \frac{1}{|\query|} \sum_{n\in \query} u_{n, k}^{(0)}$.\\
\For{$\ell = 1, 2, \ldots,$}
{
$
\displaystyle\boldsymbol{u}_n^{(\ell)} =
\operatorname{softmax}
\Big( \Big(-\frac{1}{2}(\boldsymbol{w}_k - \boldsymbol{z}_n)^\top \hat{\bs{S_k}}(\boldsymbol{w}_k - \boldsymbol{z}_n)$  \\ $\hfill+\frac{1}{2} \ln \det(\hat{\bs{S}}_k)+\frac{\lambda}{|\query|} \ln\,\pi_{k}^{(\ell)} \Big)_{ k}\Big), \; \forall n\in\query,
$

$\displaystyle  \boldsymbol{w}_k^{(\ell+1)}
= \frac{\sum_{n =1}^{N} u^{(\ell+1)}_{n, k} \boldsymbol{z}_n}{ \sum_{n =1}^{N} u^{(\ell+1)}_{n, k} }, \, \forall k\in \{1, \dots, K\},
$  

$\displaystyle\pi_{k}^{(\ell+1)} = \frac{1}{|\query|} \sum_{n\in \query} u_{n, k}^{(\ell+1)}$,  \quad $\forall k \in\{1, \dots, K\}.$
\;}
\caption{PADDLE-Cov}\label{algo:PADDLE}
\end{algorithm}
\end{center}
\section{Experiments}\label{sec:experiments}

\subsection{Experimental setting}
In our experimental setup, we leverage the pre-trained model from \cite{ciga2022self}, trained on diverse histopathological images. We structure our few-shot tasks using a sliding window of dimensions $ 5184 \times 5184$, containing mini-patches of size $1728 \times 1728$ downsampled to a resolution of $512 \times 512$. This results in query sets of 25 samples each. The support set comprises the annotated patches of the 28 train patients, and we set the penalty parameter $\lambda$ to $1250$ using validation slides.
Preprocessing includes Reinhard color normalization to mitigate staining variability \cite{946629}.

\subsection{Results}
\subsubsection{Validation on annotated test data}
Our initial evaluation focuses on the entire collection of labeled patches, which we refer to as windows, from the test set slides from 13 patients. Notably, each window is exclusively composed of mini-patches associated with a single class. In this context, we benchmark our approach against two \emph{inductive} few-shot methodologies, SimpleShot \cite{wang2019simpleshot} and Baseline \cite{chen2019closer}, which conduct inference on each mini-patch independently, as well as the state-of-the-art transductive method $\alpha$-TIM \cite{veilleux2021realistic}. In addition we provide an ablation of the terms in our classifying objective \eqref{eq:minimization_pb_histo}, evaluating the original PADDLE method (with identity covariances) and the PADDLE-Cov for $\lambda=0$. The outcomes of this comparative analysis are given in Table~\ref{table:accuracy}. Our method surpasses the other approaches, highlighting the benefits of using an appropriate Gaussian metric and of transductive inference.
\vspace{-0.1cm}
\begin{table}[ht]
  \centering
    \begin{tabular}{l|ccc}
   & Accuracy (\%) & F1-score (\%) \\ \hline
   SimpleShot \cite{wang2019simpleshot} & 48.9 & 46.4\\
   Baseline \cite{chen2019closer} & 74.4 & 72.0\\
   $\alpha$-TIM \cite{veilleux2021realistic} & 56.0  & 56.9 \\
   PADDLE \cite{martin2022towards} & 51.0 & 48.9 \\
    PADDLE-Cov ($\lambda=0$) & 77.3 & 73.8\\
    \rowcolor{lightsalmon!30} PADDLE-Cov  & \textbf{79.3} & \textbf{75.5}
    \end{tabular}
    \vspace{-0.1cm}
    \caption{Evaluation of our approach against other few-shot methods for histopathological patch classification regarding accuracy and F1-score. \label{table:accuracy}}
\end{table}
\vspace{-0.1cm}
\subsubsection{Inference on a Whole Slide Image (WSI)}
\begin{figure}[ht]
    \centering
    \includegraphics[width=0.5\textwidth]{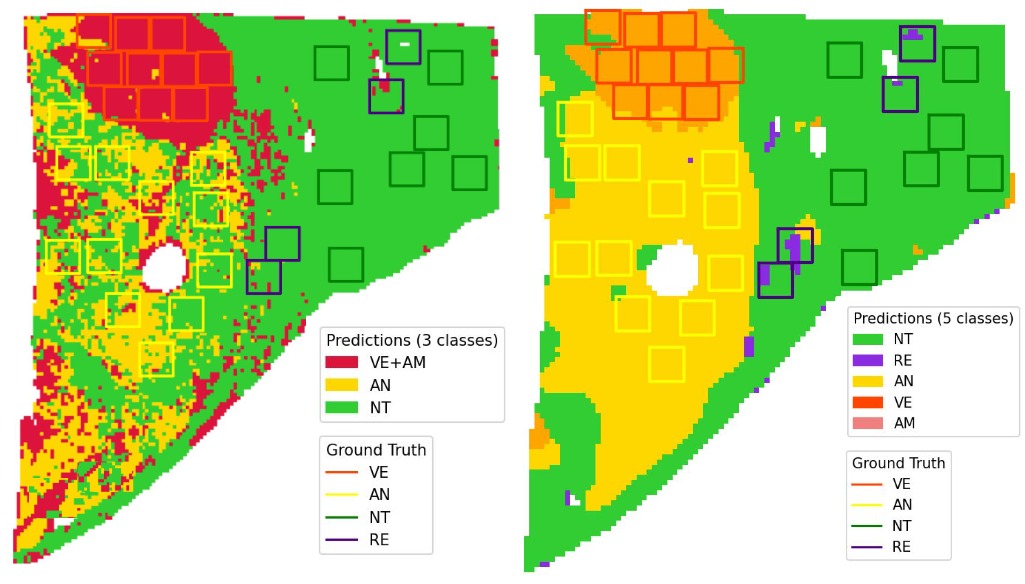}
    \caption{(left) The 3-class fully supervised model predictions. (right) The few-shot 5-class model predictions. } 
    \label{fig:WSI_preds1}
\end{figure}

\begin{figure}[ht]
    \centering
    \includegraphics[width=0.5\textwidth]{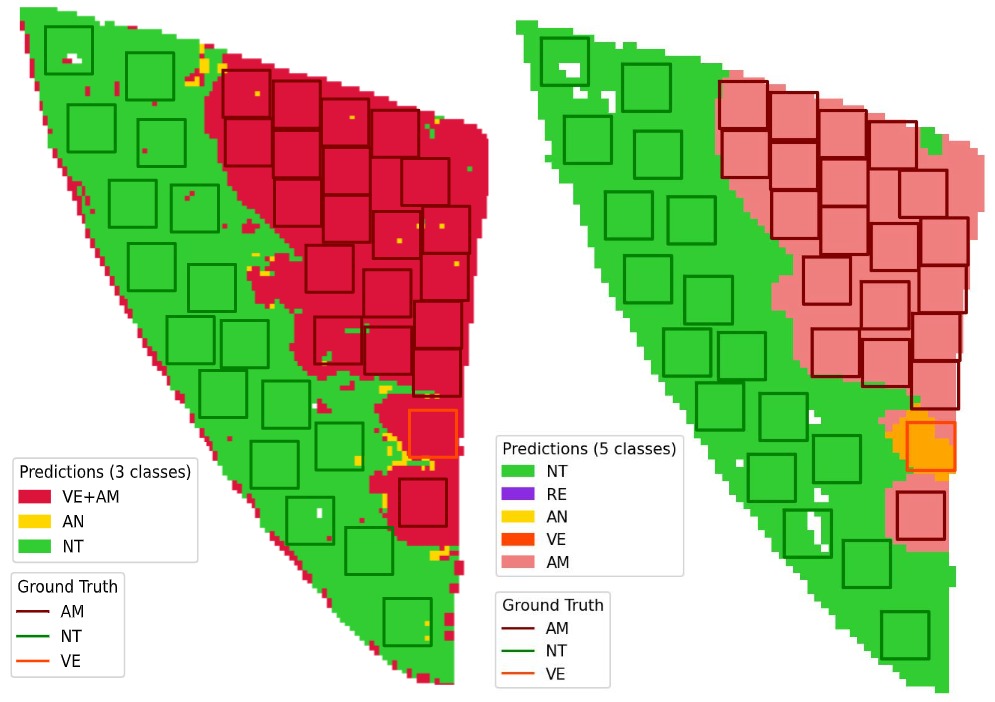}
    \caption{(left) The 3-class fully supervised model predictions. (right) The few-shot 5-class model predictions. } 
    \label{fig:WSI_preds2}
\end{figure}

In our second evaluation, we aim to compare the predictions made by our 5-class few-shot classifier trained on 28 patients with those of a 3-class fully supervised model on WSIs. The 3-class fully supervised model is a CNN based on ResNet34, which was trained using 800K patches, based on the 87 patients cohort, to classify tissues as non-tumor (NT), non-pejorative tumor (AN), or pejorative tumor (VE+AM). Creating a 5-class supervised model was hardly achievable, as two classes are notably under-represented. Figures \ref{fig:WSI_preds1} and \ref{fig:WSI_preds2} display the predictions of both models on a WSI where the colored squares represent the annotations (ground truth) made by the pathologists.\\
Both models reliably identify non-tumoral (green squares) and pejorative regions (orange/brown squares), while the conventional trabecular architecture (yellow squares) is better detected by the few-shot model on the WSI in Figure \ref{fig:WSI_preds1}.
Moreover, training the few-shot model on 5 classes enables detailed detection of the architectures, which the 3-class model can not achieve. In particular, our model accurately distinguishes the VE architecture within the pejorative regions, providing 100\% certainty in differentiating VE from AM in both WSIs. Additionally, it detected hemorrhagic regions (RE, purple squares), which were logically misclassified by the fully supervised model in Figure \ref{fig:WSI_preds1}.
Lastly, the 5-class model exhibits remarkable proficiency in defining homogeneous regions across the entire WSIs, unlike the 3-class model, which analyzes individual patches independently. The 5-class model contextual understanding allows for consideration of interdependencies between neighboring patches, leading to a more cohesive interpretation of the data.
\vspace{-0.5cm}
\vspace{-0.6cm}
\vspace{-0.55cm}

\section{Conclusion}\label{sec:conclu}
To wrap up, we have introduced an innovative transductive few-shot learning method tailored to classify histopathological images. This approach effectively overcomes significant obstacles, notably data scarcity and class imbalance. 
Our study emphasizes the adaptability and promise of our method in the domain of biomedical imaging. Its success not only emphasizes the feasibility of our approach in tackling practical challenges but also paves the way for its wider application in various medical imaging scenarios.


\vfill\pagebreak

\section{Ethics approval}
The study conformed to the General Data Protection Regulation (GDPR) and was approved by the Institutional Review Board of Mondor Hospital (IRB\#00011558) (notification number: 2022-135).

\section{Acknowledgments}
\label{sec:acknowledgments}
We would like to thank Dr Laura Claude of the Department of Pathology of the CHU de Rouen, the surgical team of the Centre H\'epato-Biliaire of the H\^opital Paul Brousse, and the technicians of the Department of Pathology of the H\^opital Bic\^etre.
\bibliographystyle{IEEEbib}
\bibliography{strings,refs}

\end{document}